\documentclass[11pt, a4paper]{article}

\usepackage{amsmath, amssymb, amsfonts}
\usepackage[utf8]{inputenc} 
\usepackage[english]{babel}
\usepackage{graphicx} 

\usepackage{subfigure} 

\usepackage{fancyhdr} 
\usepackage{color}
\definecolor{dkblue}{rgb}{0,0.1,0.5}
\definecolor{lightblue}{rgb}{0,0.5,0.5}
\definecolor{dkgreen}{rgb}{0,0.4,0}
\definecolor{dk2green}{rgb}{0.4,0,0}
\definecolor{dkviolet}{rgb}{0.6,0,0.8}

\usepackage{listings}

\newtheorem{mtheorem}{Theorem}[section]
\newtheorem{definition}[mtheorem]{Definition} 

\newtheorem{lemma}[mtheorem]{Lemma}
\newtheorem{remark}[mtheorem]{Remark}
\newtheorem{proposition}[mtheorem]{Proposition}

\DeclareMathOperator{\Rea}{Re}
\DeclareMathOperator{\Ima}{Im}
\DeclareMathOperator{\Mob}{M{o}bius}

\newcommand{\ol}{\overline}
\newcommand{\ul}{\underline}
\newcommand{\wt}{\widetilde}

\newcommand{\ssb}{{\mathcal B}}
\newcommand{\ssc}{{\mathcal C}}

\newcommand{\zz}{{\mathbb Z}}

\title{Theorem of three circles in Coq}
\author{Julianna Zsid\'o}
\date{}

\begin{document}
\maketitle

\maketitle

\begin{abstract}
The theorem of three circles in real algebraic geometry guarantees the termination and correctness of an algorithm of isolating real roots of a univariate polynomial. 
The main idea of its proof is to consider polynomials whose roots belong to a certain area of the complex plane delimited by straight lines. After applying a transformation involving inversion this area is mapped to an area delimited by circles. We provide a formalisation of this rather geometric proof in Ssreflect, an extension of the proof assistant Coq, providing versatile algebraic tools. They allow us to formalise the proof from an algebraic point of view.
\end{abstract}

\section{Introduction}
\label{intro}

The theorem of three circles that is the subject of this paper is not to be confused with the Hadamard three circle theorem in complex analysis. Our area of interest is algorithmic real algebraic geometry, for which \cite{bpr} is our main reference hereinafter. Before stating the theorem of three circles, which is called as such in \cite{bpr}, chapter 10, we first introduce some necessary vocabulary and notations.

\begin{figure}[htb!]
\begin{centering}
\includegraphics[width=8cm]{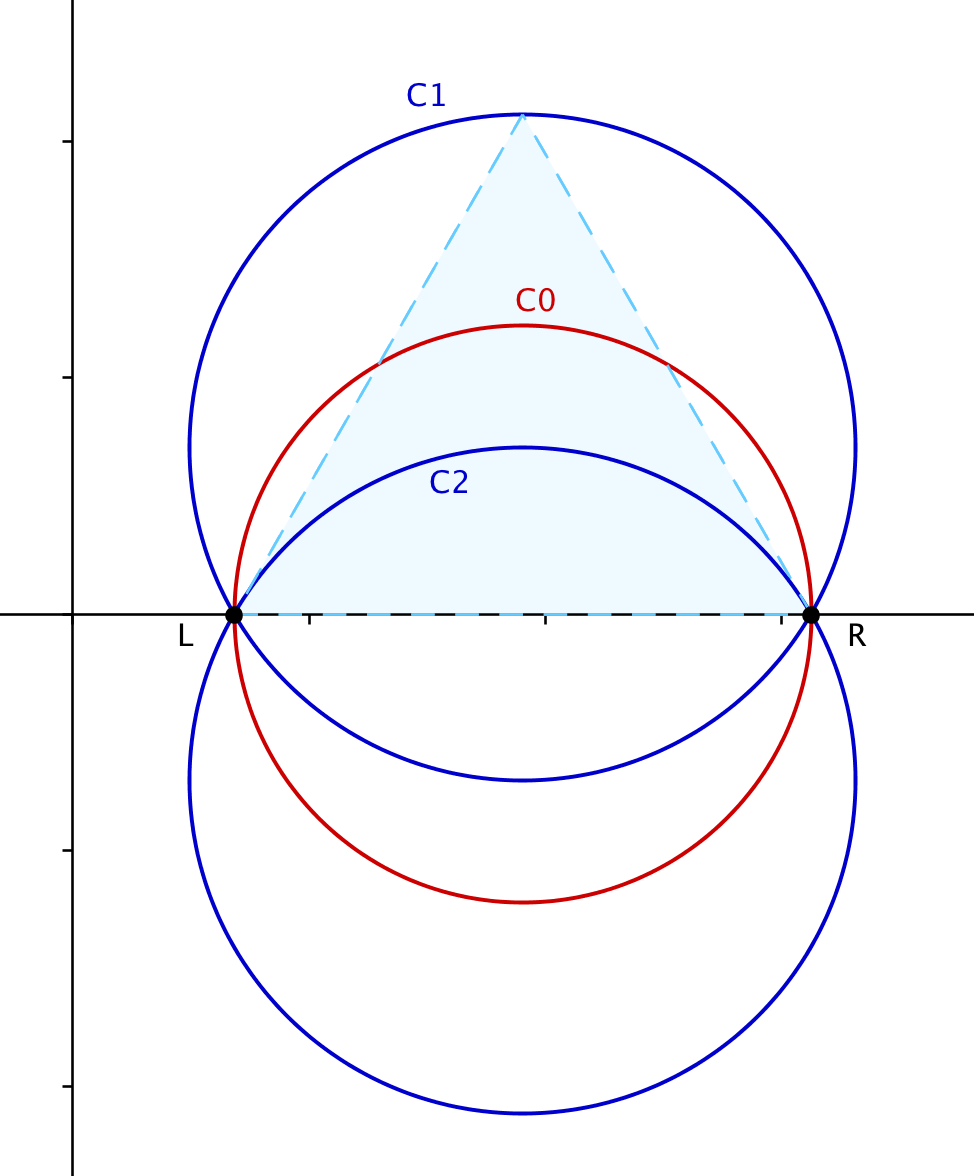}
\caption{The delimiting circles of $\ssc_0$, $\ssc_1$ and $\ssc_2$}
\label{fig_3cercles}
\end{centering}
\end{figure}

Let us fix an open real interval $(l,r)$ and consider the following three open discs of the complex plane, see figure \ref{fig_3cercles}:
\begin{itemize}
 \item $\ssc_0$ the disc bound by the circle with diameter $(l,r)$;
 \item $\ssc_1$ the disc bound by the circumcircle of the equilateral triangle with base $(l,r)$ and whose vertices have non-negative imaginary parts;
 \item $\ssc_2$ the disc symmetric to $\ssc_1$ with respect to the real axis.
\end{itemize}

\begin{figure}[htb!]
 \centering
  \begin{subfigure}
    \centering
    \includegraphics[width=5cm]{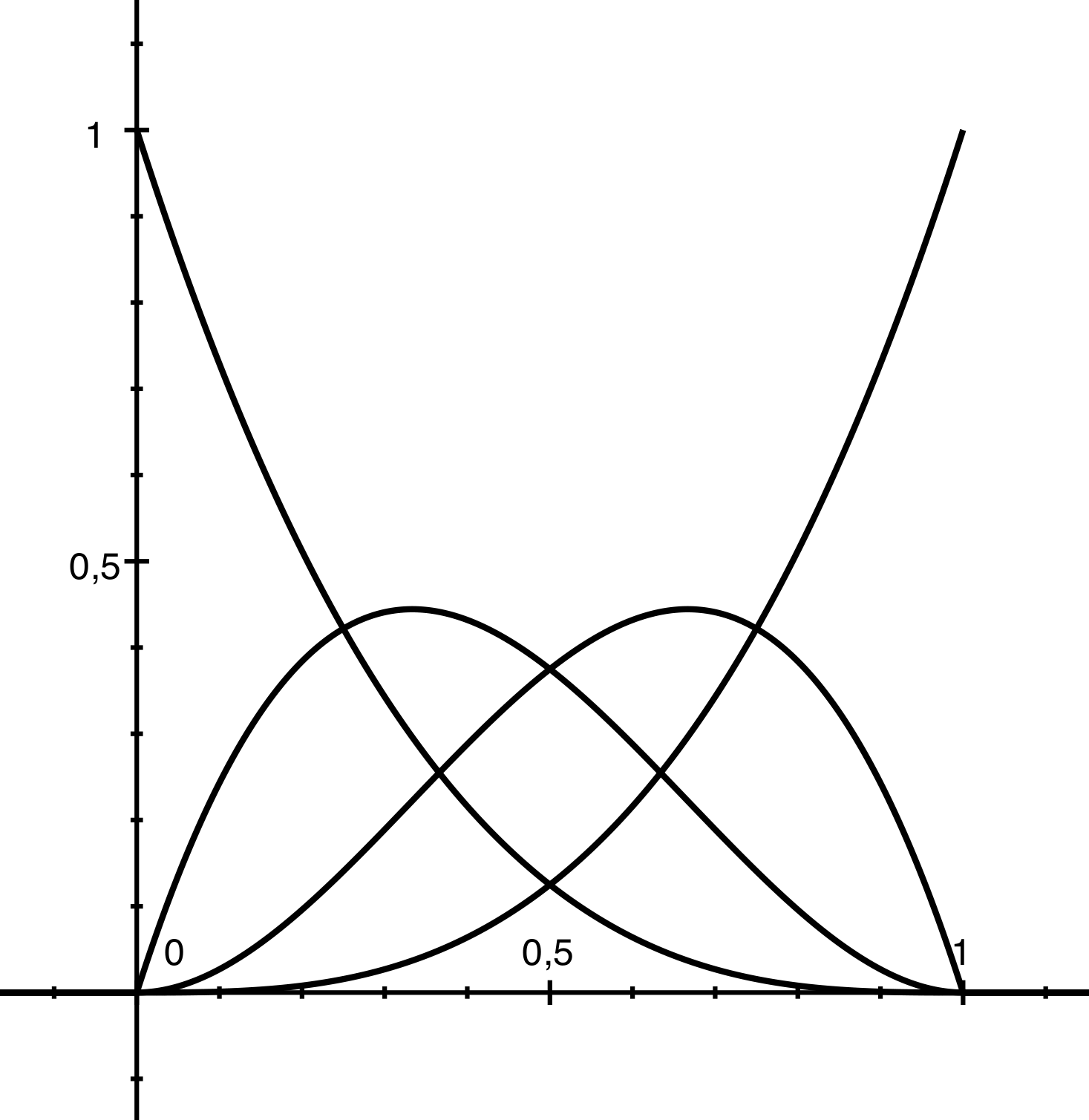}
  \end{subfigure}
  \begin{subfigure}
    \centering
    \includegraphics[width=5cm]{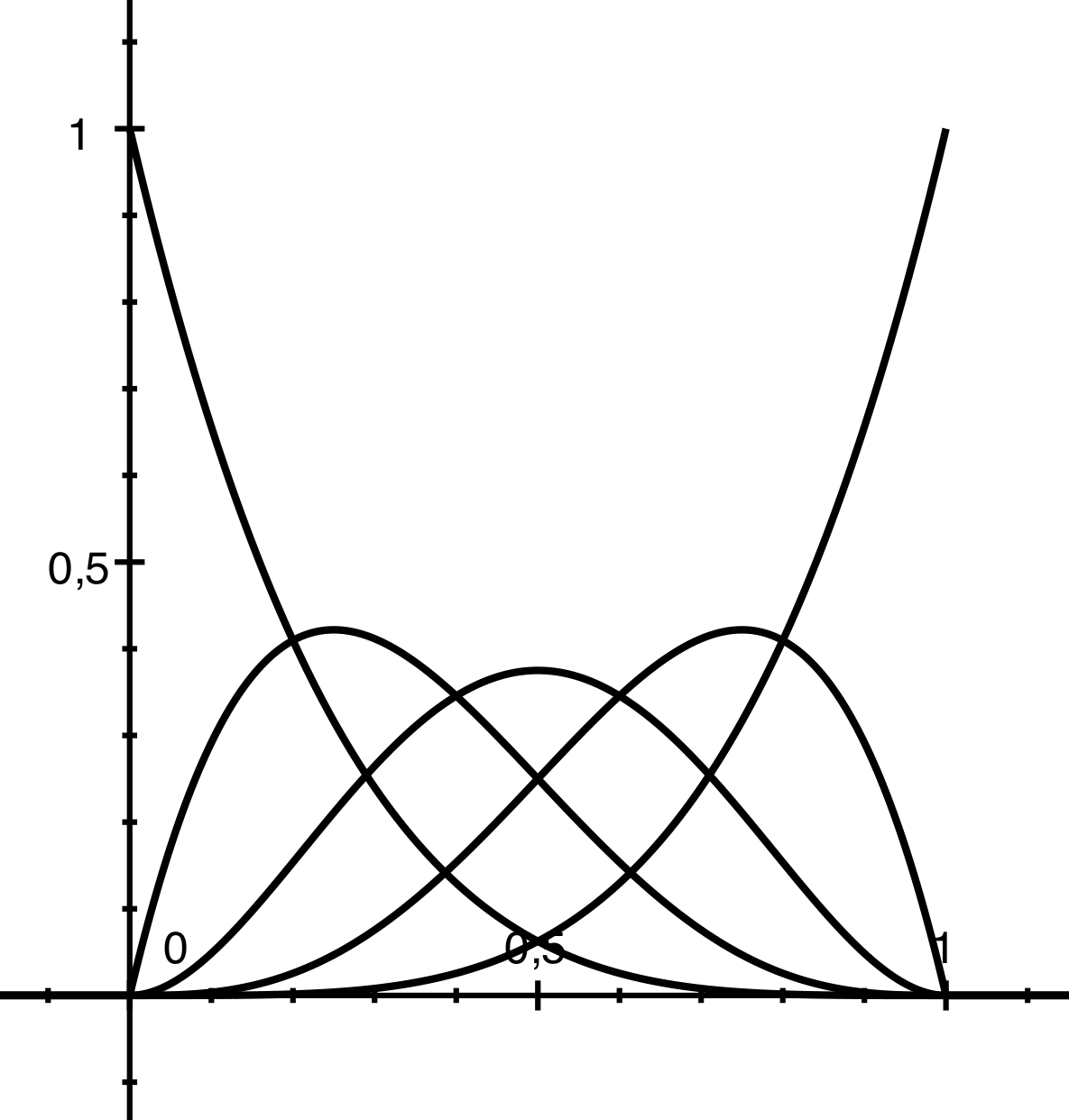}
  \end{subfigure}
  \caption{Bernstein polynomials for $n=3$ and $n=4$ on the interval $(0,1)$}
  \label{fig_bern}
\end{figure}

Next we give some intuitive elements of the theory of so--called {\em Bernstein polynomials} needed for the theorem of three circles. Bernstein polynomials are associated to a certain interval $(a,b)$, and a degree $n$, see figure \ref{fig_bern} for $(a,b) = (0,1)$, $n=3$ and $n=4$. They form a basis of $\Pi_n$, the vector space of polynomials of degree at most $n$. Bernstein polynomials can be used to approximate continuous functions on $(a,b)$. Moreover they provide the control points for Bezier curves, which play an important role in image manipulation programs for example.

The coefficients of a polynomial expressed in the Bernstein basis are its {\em Bernstein cefficients}. In figure \ref{fig_bern}, we can see that the Bernstein polynomials have maxima in different points. Given a polynomial in a Bernstein basis, intuitively speaking each Bernstein coefficient describes the behaviour of the polynomial in an interval around the maximum of the corresponding Bernstein polynomial. This does not mean that if a Bernstein coefficient is negative the polynomial is necessarily negative on the interval under its influence, but under certain circumstances it can mean this.

The statement of the theorem of three circles is the following. Let $P$ be a polynomial with real coefficients. If $P$ has no roots in $\ssc_0$, then there is no variation of signs in the sequence of Bernstein coefficients of $P$. If $P$ has exactly one simple root in the union $\ssc_1 \cup \ssc_2$, then there is exactly one variation of signs in the sequence of Bernstein coefficients of $P$. Note that the Bernstein coefficients of $P$ and the disks $\ssc_0$, $\ssc_1$ and $\ssc_2$ depend on the previously fixed reals $l$ and $r$.

This theorem is in a certain way reciprocal to Descartes' rule of signs, which states that the number of sign variations in the sequence of coefficients is an upper bound for the number of positive real roots (counted with multiplicities) and the difference of these two numbers is a multiple of 2. For the cases of sign variation 0 and 1, this rule gives the exact number of positive real roots.

\begin{figure}[htb!]
\begin{centering}
\includegraphics[width=8cm]{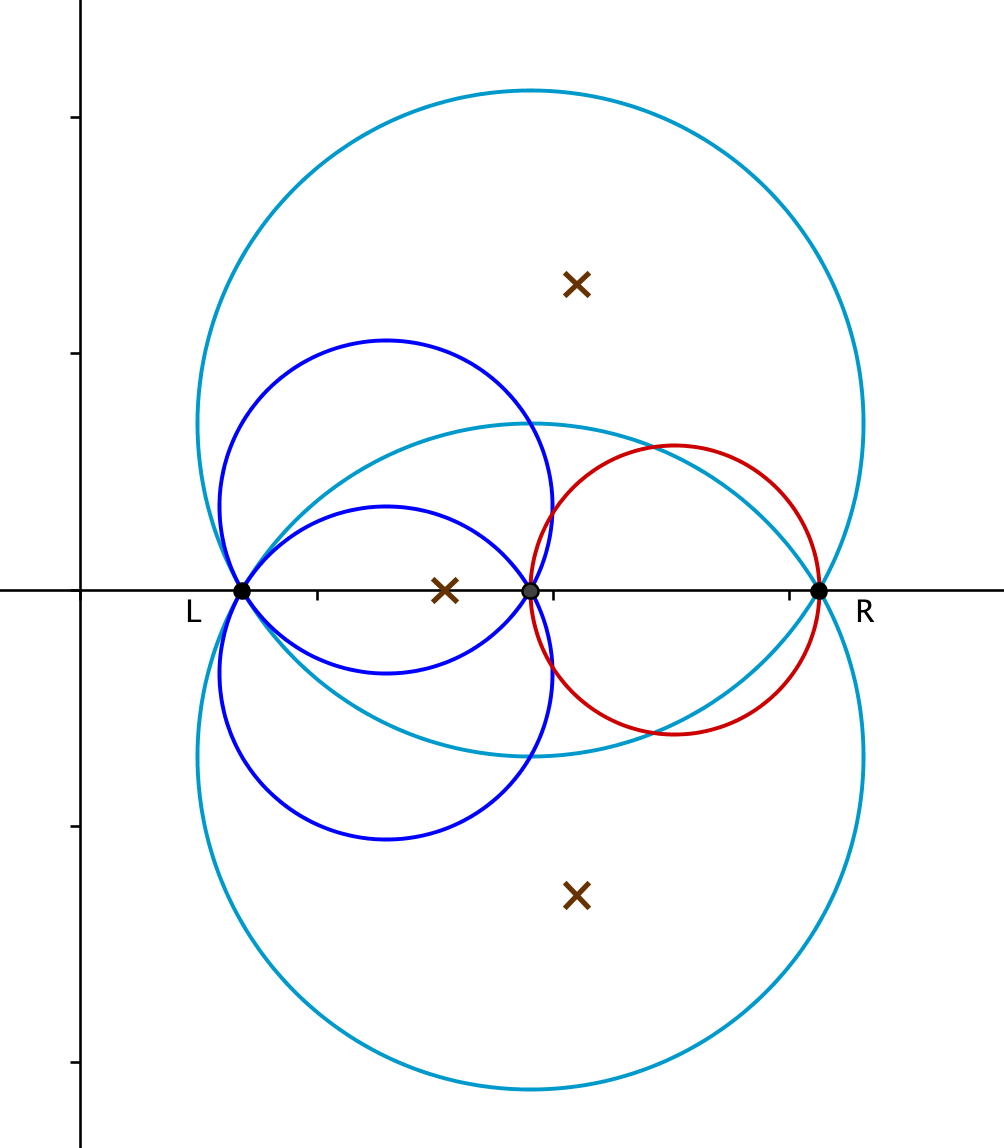}
\caption{Three roots in the union of the large discs, only one real root in the union of the two small discs on the left}
\label{fig_4cercles}
\end{centering}
\end{figure}

The theorem of three circles guarantees the correctness and termination of the algorithm for real root isolation using Descartes' method, such as presented in \cite{bpr}. One possible terminating step could be similar to the one in figure \ref{fig_4cercles}. The algorithm bisects intervals in each iteration and then checks sign variations on the intervals. If there are zero or one variations, by some arguments it concludes that there is no or one real root respectively. Otherwise it continues bisecting. The theorem of three circles says that if enough iterations are made and thus the intervals are small enough so that $\ssc_0$ contains no real root or $\ssc_1 \cup \ssc_2$ contains exactly one real root, then the algorithm will step into the terminating branch.

Bernstein polynomials occur when dealing with different mathematical problems, mainly in effective or algorithmic algebraic geometry. There are a number of recent works involving the formalization of Bernstein polynomials. The project Flyspeck \cite{flyspeck} intends to give a formal proof of the Kepler conjecture, see for example \cite{hales_etal}. This conjecture deals with sphere--packing in three dimensional (Euclidean) space, and was fomulated as such by J. Kepler in the 17th century. Its proof was given in 1998 by T. C. Hales, using exhaustively computations carried out by a computer, such as checking over a thousand nonlinear inequalities. The formalisation of this latter mentioned part was carried out in the Flyspeck project and is based on polynomial approximations using Bernstein bases. In particular one of the authors, R. Zumkeller provides a global optimisation tool based on Bernstein polynomials in Coq and in Haskell, see for example \cite{zumkeller_phd}. 
Another recent formalisation of Bernstein polynomials is the one by C. Mu\~nos and A. Narkawicz \cite{munoz_narkawicz} from NASA. They formalized an algorithm in the PVS proof assistant for finding lower and upper bounds of the minimal and maximal values of a polynomial which makes use of (multivariate) Bernstein polynomials. 
A formal study of Bernstein polynomials in the proof assistant Coq was also realized, see \cite{bernstein}. The authors of \cite{bernstein} formalised in their work the above and vaguely mentioned arguments for the conclusions: ``0(1) change of signs in the Bernstein coefficients'' implies ``no (one) real root'' (w.r.t. a fixed open interval). So together with their work, the formalisation of the theorem of three circles provides necessary pieces for the formal proof of the termination of the algorithm of real root isolation based on bisections.

Algorithms for finding and separating real roots of polynomials play an important role in computer algebra, for example in the cylindrical algebraic decomposition algorithm. Providing a formal proof of the cylindrical algebraic decomposition (CAD) algorithm has been an active field of research in the last decade, see for example works of A. Mahboubi \cite{mahboubi06,mahboubi_phd,mahboubi07}. So our interest in formalising the theorem of three circles can be regarded as part of the efforts contributing to the formalisation of the CAD.

The CAD algorithm (due to G.E. Collins, developed in the 1970's) is an algorithm of quantifier elimination in real closed fields, and it represents at the same time an effective proof of one of Tarski's results from the 1950's, namely that the theory of real closed fields is decidable. The theory of real closed fields deals with polynomial equations and inequalities and roughly speaking describes real arithmetic. Decidability means here that the CAD is an algorithm that decides whether a given sentence in the first--order language of real closed fields is provable from the axioms of real closed fields. The algorithm is interesting on the one hand in real algebraic geometry when dealing with semialgebraic sets (sets described by polynomial inequalities) and on the other hand in mathematical logic since it provides an important theoretical result on real arithmetic. The CAD algorithm represents also an improvement of Tarski's historical algorithm from another point of view. Its complexity is double exponential 
whereas Tarski's algorithm has complexity of an exponential tower in the number of quantified variables. The research for complexity improvements of the CAD algorithm is still today an active field of research.

Recent developments in fast algorithms to isolate real roots of polynomials, in particular works of M. Sagraloff \cite{sagraloff09,sagraloff11,sagraloff12}, reinforced our motivations for formalising the theorem of three circles.
The three disks $\ssc_0$, $\ssc_1$ and $\ssc_2$ represent two special cases of Obreshkoff areas. These areas are unions of open discs similar to $\ssc_1 \cup \ssc_2$, but whose center points see the open interval $(l,r)$ under the angle $\frac{2 \pi}{k+2}$ for certain positive integers $k$. Obreshkoff lenses, which are intersections of the two discs similar to $\ssc_1$ and $\ssc_2$, together with the corresponding Obreshkoff areas play again a major role in proving the correctness and termination of the NewDsc algorithm \cite{sagraloff12}, which is an algorithm based on Descartes' method with Newton style iterations. So by formalising the special cases, we provide tools for the formalisation of the Obreshkoff areas and lenses (and an analogous theorem), which for their part are necessary for the formal verification of the NewDsc algorithm for example. 

The main contribution of this work is the formalisation of the three circle theorem. 
We chose to do it with the Ssreflect extension \cite{ssreflect} (whose name is derived from small-scale reflection) of Coq \cite{coq}, since it provides versatile tools for dealing with algebraic structures and polynomials. An exhaustive introduction for Coq is for example Coq'Art, \cite{coqart} and complementary technical details can be found in the Coq manual \cite{CoqManualV84}. Moreover \cite{ssreflect-jfr} provides a nice introduction to Ssreflect. This work is partially supported by the European project ForMath \cite{formath}.

\section{Mathematical setting and prerequisites}

The theorem of three circles is valid in any real closed field $R$, not only in the field of real numbers. The complex plane is replaced by the algebraic extension $C = R[i] = R[T]/(T^2+1)$ of $R$.

Moreover there is a certain number of prerequisite results which are needed for the theorem and its proof.  

\subsection{Bernstein coefficients}
\label{bernstein}

As we already mentioned in the introduction, the assertion of the theorem involves Bernstein coefficients, which are the coefficients of a given polynomial $P$ of degree $n$ in the Bernstein basis of the vector space $\Pi_n$ of polynomials of degree at most $n$. 

The Bernstein basis of $\Pi_n$ consists of the Bernstein polynomials $B_{n,i,l,r}$, which are defined on the open interval $(l,r)$ as follows
\[B_{n,i,l,r}(X) =  {{n} \choose {i} }\frac{(X-l)^i (r-X)^{n-i}}{(r-l)^n} \]
for $i=0,\ldots,n$.

The Bernstein coefficients of a polynomial $P$ can be computed from the coefficients of another polynomial $Q$ which is obtained by applying a certain number of polynomial transformations on $P$. Before stating the corresponding proposition, we introduce the necessary transformations:

\begin{enumerate}
 \item Translation by $c \in R$: $T_c (P (X)) = P(X-c)$, 
 \item Scaling by $c \in R$: $S_c(P(X)) = P(cX)$,
 \item Inversion: $I_n(P(X)) = X^n P(1/X)$.
\end{enumerate}

\begin{proposition}[\cite{bpr}]
Let $P(X) = \sum_{i=0}^n b_i B_{n,i,l,r} (X)$ be a polynomial of $R[X]$ of degree at most $n$ and let $Q(X) = T_{-1} (I_n (S_{r-l} (T_{-l}(P(X)))))$ whose coefficients (in the monomial basis) are denoted by $c_i$.
Then $c_{n-i} = {n \choose i} b_i$.
\label{bernQ}
\end{proposition}

The proof of this proposition consists of the computation of the coefficients of $Q$.

\begin{definition}
Inspired by \cite{eigenwillig_phd}, we call the sequence of the above four transformations of the polynomial $P$ a {\em M\"obius transformation} of $P$; we will write $\Mob(P)$ for the polynomial $Q$ and call it the {\em M\"obius transform} of $P$.
\end{definition}

\subsection{Normal polynomials}
\label{normal_math}

Another ingredient in the proof of the theorem relies on results about so--called {\em normal} polynomials. A polynomial $P (X) =\sum_{i=0}^p a_i X^i$ is normal if and only if it satisfies the following conditions:
\begin{enumerate}
 \item $0 \leq a_i$ for all $i=0,\ldots,p$;
 \item $0 < a_p$;
 \item $a_{i-1} a_{i+1} \leq a_i^2$ for all $i=0,\ldots,p$ (where coefficients with indices out of range are equal to 0);
 \item for all $j \in \{0,\ldots,p\}$ such that $0 < a_j$ then $0< a_i$ for all $i=j,\ldots,p$.
\end{enumerate}
So we deal with polynomials whose sequence of coefficients consists of a certain number of zeros followed by positive ones up to the leading coefficient.

We have the following (almost immediate) consequences :

\begin{lemma}[\cite{bpr}]
The polynomial $X-x$ is normal if and only if $x \leq 0$.
\label{lemma241}
\end{lemma}

\begin{lemma}[\cite{bpr}]
A second degree polynomial with a pair of complex conjugate roots is normal if and only if its roots are contained in the area $\ssb = \{a +bi \in R[i] \phantom{i} | \phantom{i} a \leq 0, b^2 \leq 3a^2 \}$.
\label{lemma242}
\end{lemma}

\begin{lemma}[\cite{bpr}]
The product of two  normal polynomials is normal.
\label{lemma243}
\end{lemma}

The proofs of the first two lemmas are mainly computations. In the proof of the last one, one has to deal with double sums which are the coefficients of the product polynomial. It requires a tricky partition of the range of indices (or simply of $\zz^2$), the remaining part is technical but without any further difficulty.

Let us recall a definition:

\begin{definition}[\cite{bpr}]
A polynomial is called {\em monic} if and only if its leading coefficient is 1.
\end{definition}

Now with the three previous lemmas one can show the following proposition:

\begin{proposition}[\cite{bpr}]
Let $P(X) \in R[X]$ be a monic polynomial. If all its roots belong to $\ssb= \{a +bi \in R[i] \phantom{i} | \phantom{i} a \leq 0, b^2 \leq 3a^2 \}$, then $P$ is normal.
\label{prop240}
\end{proposition}

\begin{remark}
Without loss of generality one can consider only normal polynomials whose sequence of coefficients does not contain zeros. This is equivalent to considering only normal polynomials such that zero is not a root, since the multiplicity of the root in zero corresponds to the number of zero coefficients in the beginning of the sequence of coefficients.
\label{remarkno0}
\end{remark}
The main result involving normal polynomials is the following:

\begin{proposition}[\cite{bpr}]
Let $P(X) \in R[X]$ be a normal polynomial and $0 < a$, then the number of sign variations in the sequence  of coefficients of $P(X)(X-a)$ is exactly 1.
\label{prop244}
\end{proposition}

\textit{Proof sketch.}
If we denote the coefficients of $P$ by $p_i$ and its degree by $n$, then we have
\[P(X)(X-a) =  -p_0a +p_1 \biggl(\frac{p_0}{p_1} - a \biggr) X + \ldots + p_n \biggl(\frac{p_{n-1}}{p_n} - a \biggr) X^n + p_n X^{n+1}\]
We have $-p_0a < 0$ and $0 < p_n$, moreover the following chain of inequalities holds
\[\biggl( \frac{p_{k-1}}{p_k} - a\biggr) \leq \biggl( \frac{p_{k}}{p_{k+1}} - a\biggr)\]
for $k=1,\ldots,n-1$, because of the condition $p_{k-1} p_{k+1} \leq p_k^2$ for the normal polynomial $P$ and the fact that all $p_k$ are positive.

So the sequence of coefficients of $P(X)(X-a)$ without the first and the last elements has at most one sign change. If it has exactly one, $p_1 \bigl(\frac{p_0}{p_1} - a \bigr) \leq 0$ and $0 \leq p_n \bigl(\frac{p_{n-1}}{p_n} - a \bigr)$, so there is no sign change between the first and second or between the last and before last coefficients. If there is no sign change in the middle coefficients, $p_1 \bigl(\frac{p_0}{p_1} - a \bigr)$ and $p_n \bigl(\frac{p_{n-1}}{p_n} - a \bigr)$ have the same sign. If they are both negative, there is a sign change from $p_n \bigl(\frac{p_{n-1}}{p_n} - a \bigr)$ to $p_n$, if they are both positive, there is a sign change from $-p_0 a$ to $p_1 \bigl(\frac{p_0}{p_1} - a \bigr)$.

\section{Using existing theories of Coq}

In this section we are going to give some details on the previously formalised theories in Coq on which we base our proof.

But first let us point out that the notation for functions $x \mapsto f(x)$ is written as \lstinline!fun x => f x! in Coq.

We base our proof on the tools provided by the standard libraries of Ssreflect, developed by the Mathematical Components Project \cite{ssreflect}, such as \textsf{ssreflect}, \textsf{ssrbool}, \textsf{ssrnat}, \textsf{seq}, \textsf{path}, \textsf{poly}, \textsf{polydiv}, \textsf{ssralg} and \textsf{ssrnum}. The latter two contain a hierarchy of algebraic structures, such as groups, rings, integral domains, fields, algebric fields, closed fields and their ordered counterparts. An exhaustive explanation of the organisation and formalisation of these structures can be found in section 2 of \cite{cm_ssr} or in chapters 2 and 4 of \cite{cohen_phd}.

Moreover we use less standard libraries, such as \textsf{complex}, \textsf{polyorder}, \textsf{polyrcf}, \textsf{qe\_rcf\_th}, \textsf{pol} and \textsf{bern}. We are going to explain the provided elements and notations of these libraries which are necessary to understand the codes shown in the next section.

In the following \lstinline{R} denotes a real closed field, unless stated otherwise, and \lstinline!C = complex R!  its algebraic extension. The real part of a complex number \lstinline?z? is denoted by \lstinline!Re z! and its imaginary part by \lstinline!Im z!.

The type \lstinline!{poly R}! is provided for a ring \lstinline!R!, representing the type of univariate polynomials with coefficients in \lstinline!R!.  The indeterminate $X$ is written \lstinline!'X!, the $k$--th coefficient of the polynomial \lstinline!p! is written \lstinline!p`_k!, the composition of two polynomials \lstinline!p! and \lstinline!q! is written \lstinline!p \Po q! and the degree of \lstinline!p! is given by \lstinline!(size p).-1!. The leading coefficient of \lstinline!p! is written \lstinline!lead_coef p!. We are using the predicate \lstinline!root p x! which is true iff $p(x) = 0$, i.e. $x$ is a root of $p$. Moreover \lstinline!p \is monic! represents the predicate monic, so this expression is true iff the leading coefficient of \lstinline!p! is equal to 1.

Polynomials are identified with the sequence of their coefficients. So indirectly, but often even directly, we deal with sequences when dealing with polynomials. The length of a sequence \lstinline!s! is called \lstinline!size s! and its $i$-th item is written \lstinline!s`_i!. We are going to deal with the \lstinline!all! and \lstinline!sorted! constructions. The expression \lstinline!all a s! is true iff the predicate \lstinline!a! holds for each item of the sequence \lstinline!s!. The expression \lstinline!sorted a s! is true iff the binary relation \lstinline!a! is true for each pair of consecutive items of \lstinline!s!. Moreover we make some use of \lstinline!drop! and \lstinline!take! which are transformations of sequences, allowing to drop a certain number of items from the beginning of the sequence and to take a certain number of items starting from the beginning of the sequence (respectively). The filter operation on sequences comes handy too, \lstinline!filter a s! is the sequence consisting of 
the items of \lstinline!s! which satisfy the predicate \lstinline!a!, its notation is \lstinline![seq x <- s | (a x) ]!, where the predicate is of the form \lstinline!fun x => a x!. The mask operation is quite similar to \lstinline!filter!, but it takes a boolean sequence \lstinline!b! and another sequence \lstinline!s! as inputs and returns the list consisting of items of \lstinline!s! with all indices $i$ for which \lstinline!b`_i = true! holds.  The operation \lstinline!zip! takes two sequences as input and returns the sequence consisting of pairs of items, the first item in a pair from the first sequence, the second from the second sequence. The length of the ``zipped list" is the length of the shorter input list. Mapping lists is done via \lstinline!map!, it applies a given map point-wise to the items of the sequence. The notation for maps of lists is \lstinline![seq (f x) | x <- s]!, where we apply \lstinline!fun x => f x! to the items of \lstinline!s!.

To talk about the number of sign changes in a sequence of elements of a real closed field, we use the function \lstinline!changes!: 
\begin{lstlisting}
fun R : rcfType =>
fix changes (s : seq R) : nat :=
  match s with
  | nil => 0
  | a :: q => ((a * head q < 0) + changes q)
  end.
\end{lstlisting}
by interpreting true as 1 and false as 0. This function cannot be used directly because it computes (for our purpose) erroneous values in the presence of 0 coefficients. It adds 1 to the count if $ab < 0$, writing $[a,b,\ldots]$ for the sequence $s$. So for example \lstinline!changes [ :: -1; 0; 1]! would yield 0 since $0 < 0$ is false. But in fact there is one sign change in the sequence $[-1,0,1]$, so we are going to use the function \lstinline!changes! in combination with a filter that filters out the zeros from the sequence:
\begin{lstlisting}
Definition seqn0 (s : seq R) := [seq x <- s | x $\neq$ 0]. 
\end{lstlisting}
Indeed \lstinline!changes (seqn0 [::-1;0;1])! yields 1. This definition of \lstinline!seqn0! as such is not provided by a library but it is rather a definition needed for our purposes and that complements \lstinline!changes!.

A formal study of Bernstein coefficients has already been implemented, see \cite{bernstein}, the three transformations on polynomials and the M\"obius transformation are provided by: 
\begin{enumerate}
\item Translation by $-c$:
\begin{lstlisting}
Definition shift_poly (R : ringType) (c : R) (p : {poly R}) :=
   p \Po ('X + c).
\end{lstlisting}
And its notation:
\begin{lstlisting}
Notation "p \shift c" := (shift_poly c p) (at level 50) : ring_scope.
\end{lstlisting}
\item Scaling by $c$:
\begin{lstlisting}
Definition scaleX_poly (R : ringType) (c : R) (p : {poly R}) :=
   p \Po ('X * c).
\end{lstlisting}
And its notation:
\begin{lstlisting}
Notation "p \scale c" := (scaleX_poly c p) (at level 50) : ring_scope.
\end{lstlisting}
\item Inversion:
\begin{lstlisting}
Definition reciprocal_pol (R : ringType) (p : {poly R}) :=
   \poly_(i < size p) p`_(size p - i.+1).
\end{lstlisting}
\item The M\"obius transformation of $P(X)$ (from Proposition \ref{bernQ}):
\begin{lstlisting}
Definition Mobius (R : ringType) (p : {poly R}) (a b : R) :=
   reciprocal_pol ((p \shift a) \scale (b - a)) \shift 1.
\end{lstlisting}
\end{enumerate}

\section{The proof of the theorem of three circles}

First of all let us state the theorem of three circles explicitly.

\begin{mtheorem}[\cite{bpr}]
Let $R$ be a real closed field, $l, r \in R$ s.t. $l < r$ and $P \in R[X]$ of degree $n$.
Let us write $\ssc_0$, $\ssc_1$ and $\ssc_2$ for the discs introduced in section \ref{intro}, keeping in mind the fact that the discs depend on the chosen elements $l$ and $r$.
Moreover let us write $b_P$ for the sequence of Bernstein coefficients of $P$ with respect to the Bernstein basis $\{B_{n,i,l,r}\}_{i=0,\ldots,n}$.
\begin{enumerate}
 \item If $P$ has no root in $\ssc_0$, then there is no sign variation in $b_P$.
 \item If $P$ has exactly one simple root in $\ssc_1 \cup \ssc_2$, then there is exactly one sign variation in $b_P$.
\end{enumerate}
\end{mtheorem}

Its proof can be divided into two parts, since it actually consists of two assertions: the first one dealing with $\ssc_0$ and the second one with $\ssc_1 \cup \ssc_2$.

\begin{remark}
Considering proposition \ref{bernQ}, the number of sign changes in the sequence of Bernstein coefficients of a polynomial $P$ is equal to the number of sign changes in the sequence of coefficients of the M\"obius transformation $\Mob(P)$, since reversing the sequence and multiplying each element by a positive number does not affect the number of sign changes.
\label{obda2}
\end{remark}

\begin{figure}[htb!]
 \centering
  \begin{subfigure}
    \centering
    \includegraphics[width=5cm]{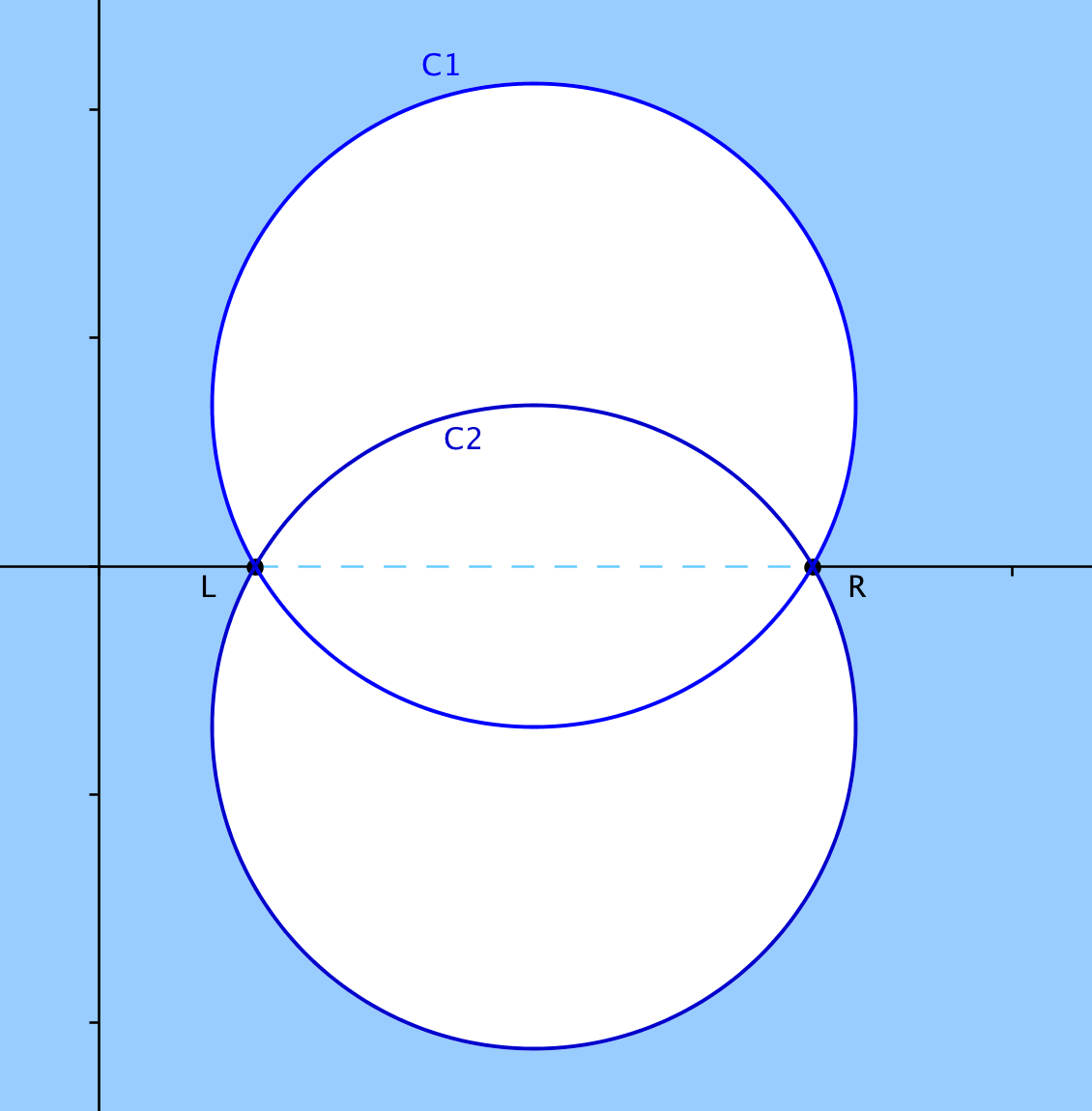}
  \end{subfigure}
  \qquad
  \begin{subfigure}
    \centering
    \includegraphics[width=5cm]{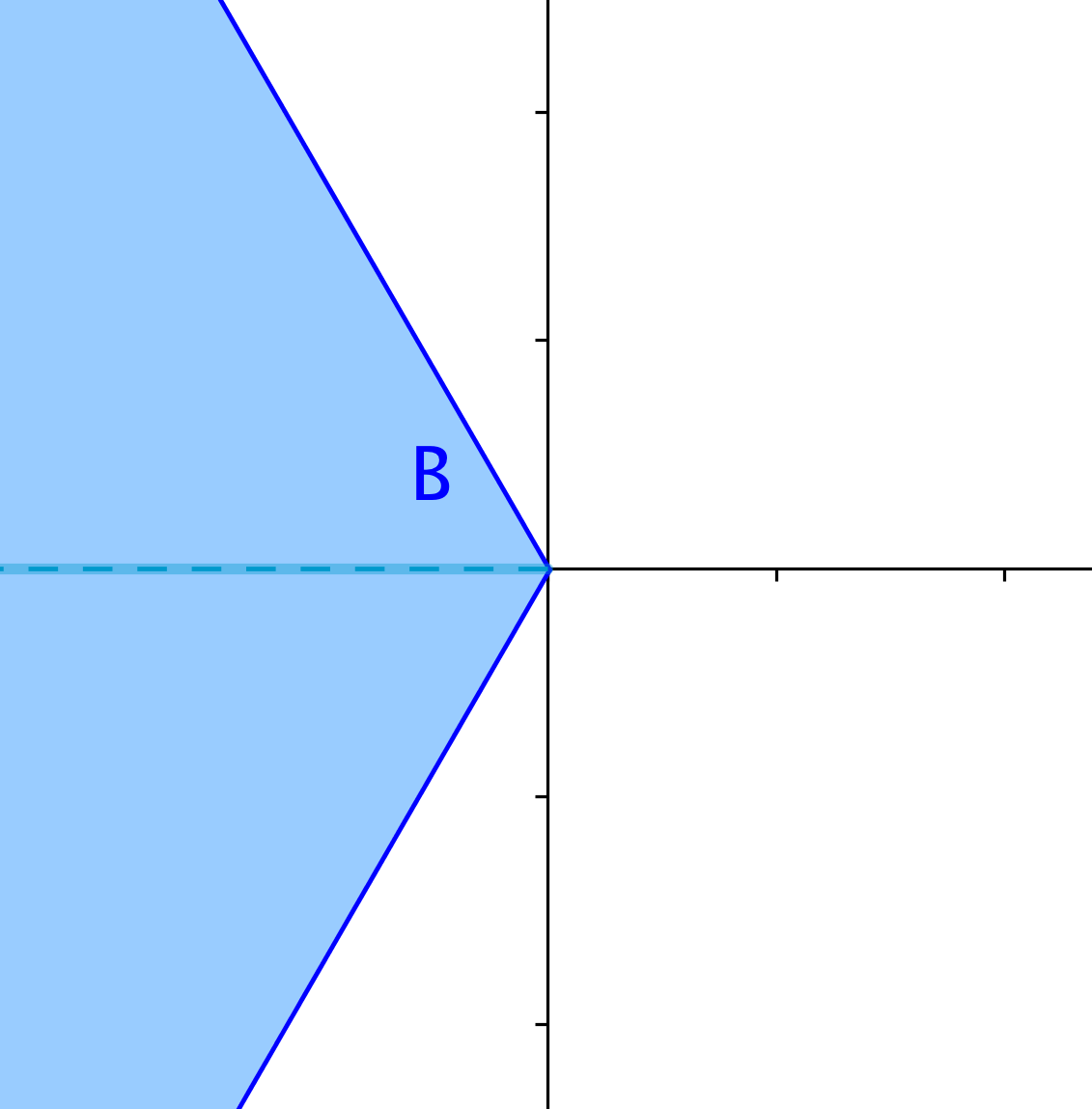}
  \end{subfigure}
  \caption{The M\"obius transformation maps the exterior of $\ssc_1 \cup \ssc_2$ to the area $\ssb$}
  \label{fig_C12}
\end{figure}

We point out the similar patterns in the proofs of the two parts. These similarities will be useful for the generalisation of the theorem we sketch in section \ref{future}.
\begin{enumerate}
 \item First we have a statement about the number of sign changes in the sequence of the coefficients of a polynomial, whose roots belong to a certain area of the complex plane. This area is the half-plane of numbers with non-positive real parts in the first part and the corresponding statement is lemma \ref{lemma239}.  The area in the second part is $\ssb$ of lemma \ref{lemma242} and of proposition \ref{prop240} and the statement about sign changes is proposition \ref{prop244}.
 \item Then by the M\"obius transformation of proposition \ref{bernQ}, this area is transformed to another area of the complex plane: the exterior of the disc $\ssc_0$ in the first part and the exterior of $\ssc_1 \cup \ssc_2$ in the second part. See figures \ref{fig_C0} and \ref{fig_C12} for these transformations.
\end{enumerate}

In the following subsections we are going to detail the proof of the theorem of three circles in three parts: \begin{itemize}
 \item Subsection \ref{first_coq} concerns the first assertion of the theorem involving $\ssc_0$, it contains both parts: before and after the M\"obius transformation.
 \item Subsection \ref{normal_coq} concerns the formalisation of the theory of normal polynomials as described in section \ref{normal_math}. The goal of this section is to prove a statement about sign changes before the M\"obius transformation.
 \item Finally subsection \ref{second_coq} concerns the second assertion of the theorem involving $\ssc_1 \cup \ssc_2$. This part of the proof uses the theory of normal polynomials and takes place after the M\"obius transformation.
\end{itemize}

\subsection{First part of the proof: using polynomials with non-negative coefficients}
\label{first_coq}

We divide this section in two subsections: before and after the M\"obius transformation.

\subsubsection{Before the M\"obius transformation}

The goal of this section is to prove the following lemma.
\begin{lemma}
Let $P(X) \in R[X]$ be a monic polynomial. If all the roots of $P$ have non-positive real parts, then there is no sign change in the sequence of coefficients of $P$.
\label{lemma239}
\end{lemma}

\textit{Proof sketch.} This is actually a simple fact, that can be proven by induction on the degree of $P$. 

If $P$ has no roots, that is, if $P$ is a constant polynomial, the assertion is obviously true.
Let $z$ be a root, so we can factor $P(X) = (X - z) P_1(X)$.

If $z$ is real, then $z \leq 0$ by hypothesis, so $0 \leq -z$ and thus by multiplying out $(X - z) P_1(X)$, one obtains only non-negative coefficients. The sequence of only non-negative coefficients does not contain any sign changes.

If $z$ has complex part different from 0, then $\ol{z}$ is a root as well and we can factor $P(X) = (X - z)(X-\ol{z}) P_2(X) = (X^2 -2 \Rea(z)X + \Rea^2(z) + \Ima^2(z)) P_2(X)$. Since $\Rea(z) \leq 0$, one obtains again only non-negative coefficients when multiplying out $(X^2 -2 \Rea(z)X + \Rea^2(z) + \Ima^2(z)) P_2(X)$.

In order to formalise this lemma, we first define a predicate \lstinline!nonneg! using the provided predicate \lstinline!all!; \lstinline!nonneg! is true iff all items of a given sequence are non-negative:
\begin{lstlisting}
Definition nonneg (s : seq R) := all (fun x => 0 $\leq$ x) s.
\end{lstlisting}
Then we prove the following lemmas representing elements of the above mentioned proof:
\begin{lstlisting}
Lemma nonneg_poly_deg1 : forall (a : R),
   nonneg ('X - a) = (a $\leq$ 0).
\end{lstlisting}

\begin{lstlisting}
Lemma nonneg_poly_deg2 : forall (z : C),
   nonneg ('X^2 - 2 (Re z) * 'X + (Re z)^2 + (Im z)^2) = ((Re z) $\leq$ 0).
\end{lstlisting}

\begin{lstlisting}
Lemma nonneg_mulr : forall (p q : {poly R}),
   (nonneg p) ->
   (nonneg q) ->
   nonneg (p * q).
\end{lstlisting}

\begin{lstlisting}
Lemma nonneg_root_nonpos : forall (p : {poly R}),
   (p \is monic) ->
   (forall z : C, root p z -> Re z $\leq$ 0) ->
   nonneg p.
\end{lstlisting}
The proof of the last lemma is done by induction on the degree of \lstinline!p! as in the proof sketch of Lemma \ref{lemma239}. This formal proof contains the largest part of the sketched proof. The remaining two lemmas are almost immediate:
\begin{lstlisting}
Lemma nonneg_changes0 : forall (s : seq R),
   (nonneg s) ->
   changes s = 0.
\end{lstlisting}

\begin{lstlisting}
Lemma monic_roots_changes_eq0 : forall (p : {poly R}),
   (p \is monic) ->
   (forall z : C, root p z -> Re z $\leq$ 0) ->
   changes p = 0.
\end{lstlisting}

Having formalised the proof of lemma \ref{lemma239}, we can now turn to the actual assertion of the theorem involving $\ssc_0$.

\subsubsection{After the M\"obius transformation}

\begin{figure}[htb!]
 \centering
  \begin{subfigure}
    \centering
    \includegraphics[width=5cm]{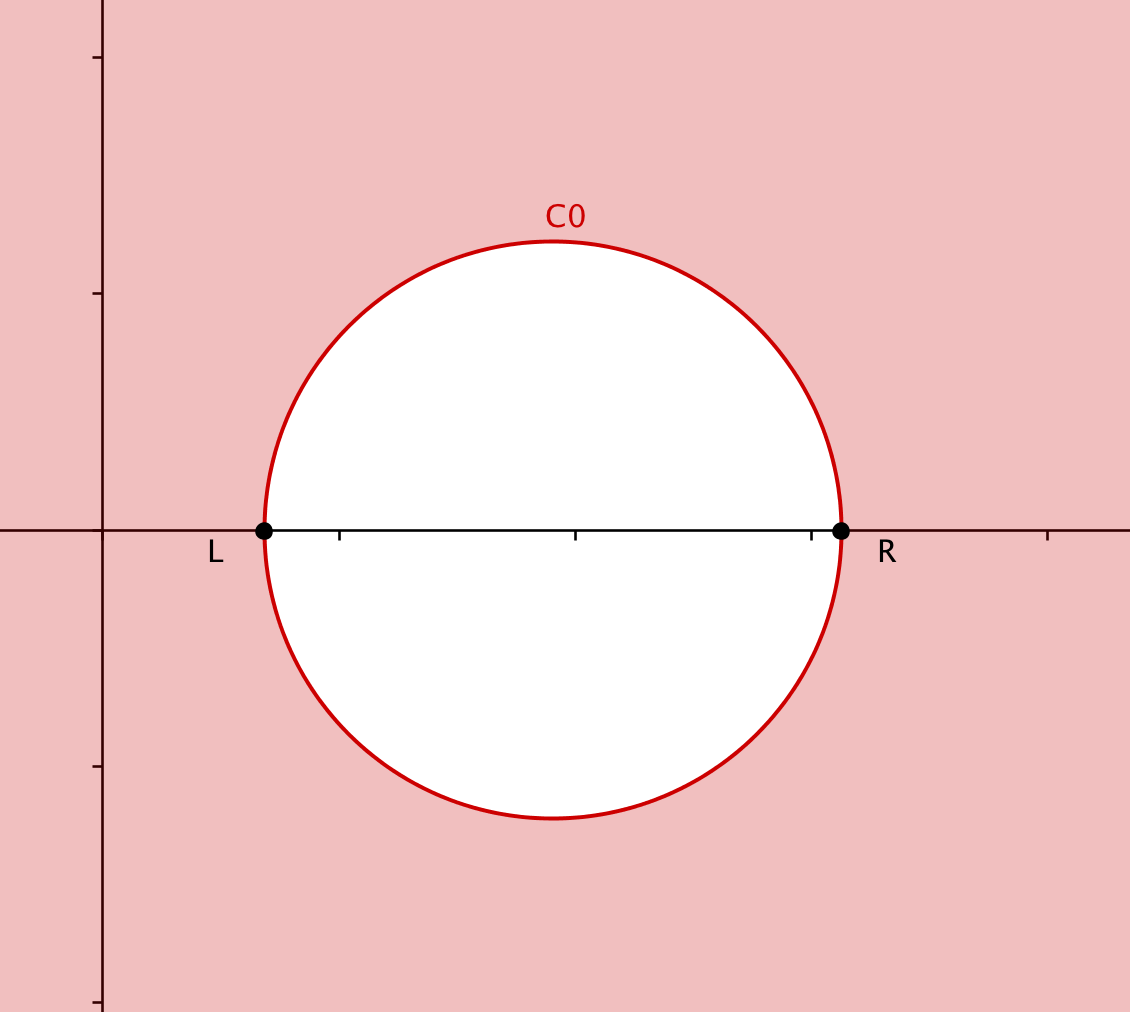}
  \end{subfigure}
  \qquad
  \begin{subfigure}
    \centering
    \includegraphics[width=5cm]{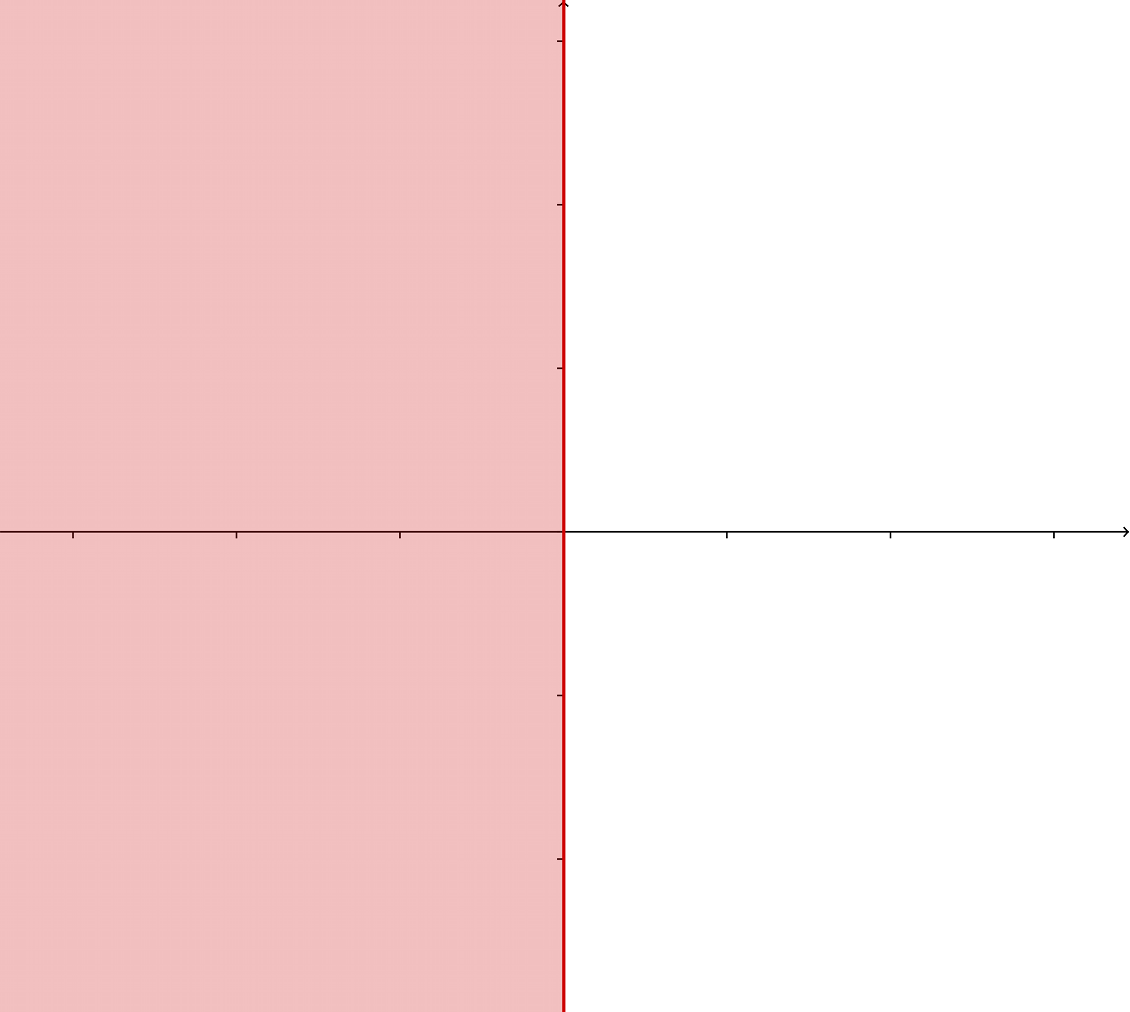}
  \end{subfigure}
  \caption{The M\"obius transformation maps the exterior of $\ssc_0$ to the half-plane with non-positive real parts}
  \label{fig_C0}
\end{figure}

Explicitly the disc $\ssc_0$ is the area of $C=R[i]$ given by 
\[\ssc_0 = \biggl\{x + yi \in R[i] \phantom{i} | \phantom{i} \Bigl(x - \frac{l+r}{2} \Bigr)^2 + y^2 < \Bigl(\frac{r-l}{2} \Bigr)^2 \biggr\},\]
or equivalently by
\[\ssc_0 = \Bigl\{x + yi \in R[i] \phantom{i} | \phantom{i} x^2 - (l + r)x + y^2 + rl < 0\Bigr\}.\]
The condition on the roots is that they are in the complementary of $\ssc_0$, so we formalise directly the predicate:
\begin{lstlisting}
Definition notinC (z : C) :=
   0 $\leq$ (Re z)^2 - (l + r) * (Re z) + (Im z)^2 + r * l.
\end{lstlisting}
By using remark \ref{obda2}, the first part of the assertion is formalised as follows:
\begin{lstlisting}
Theorem three_circles_1 : forall (p : {poly R}),
   (forall z : C, root p z -> notinC z) ->
   changes (Mobius p l r) = 0.
\end{lstlisting}
As mentioned before, we want to use lemma \ref{lemma239} in this proof. In order to apply it, we have to make sure, that the polynomial we consider is monic. The polynomial \lstinline!(Mobius p l r)! is in general not monic, but we can multiply it with the inverse of its leading coefficient and this operation does not affect the sign changes. This fact is formalised by:
\begin{lstlisting}
Lemma changes_mulC : forall (p : {poly R}) (a : R),
   (a $\neq$ 0) ->
   changes p = changes (a * p).
\end{lstlisting}
So now we can apply lemma \lstinline!monic_roots_changes_eq0!, and it remains to prove that the roots of \lstinline!(Mobius p l r)! all have non-positive real parts. Keeping in mind that this latter polynomial is in fact $\Mob(P(X)) =$\\ $T_{-1} (I_p (S_{r-l} (T_{-l}(P(X)))))$ whose roots all have non-positive real parts iff the roots of $P$ are in the complement of $\ssc_0$. When keeping track of the roots during the four transformations, this is what happens: when translating by $-l$, and then scaling by $r-l$, the roots are ``shifted" into the complement of the circle with diameter $(0,1)$. By the following inversion, the complement of the circle is mapped on the half--plane with real parts $\leq 1$ and by translating this area by $-1$, we obtain the half--plane of numbers with non-positive real parts. See figure \ref{fig_C0}.

This equivalence is not formalised as such, but is represented by the two lemmas:
\begin{lstlisting}
Lemma root_Mobius_C_2 : forall (p : {poly R}) (z : C) (l r : R),
   (z + 1 != 0) ->
   root p ((r + l * z) / (z + 1)) = root (Mobius p l r) z.
\end{lstlisting}
and
\begin{lstlisting}
Lemma notinC_Re_lt0_2 : forall (z : C), (z + 1 != 0) ->
   (notinC ((r + l * z) / (z + 1))) = (Re z <= 0).
\end{lstlisting}

One can see that when applying the above lemmas in the formal proof, the case of a root in $z = -1$ is excluded. In fact this case is treated separately, the assertion in this case can be shown directly since the real part of $-1$ is non-positive. This completes the formal proof of the first part of the theorem of three circles.

We see that the proof can be formalised basically the same way as the proof on paper suggests, thanks to all the existing theories and tools provided by the Ssreflect library.

\subsection{Second part of the proof: formalising normal polynomials}
\label{normal_coq}

This subsection contains the formalisation of subsection \ref{normal_math} about normal polynomials.
The goal of this subsection  is to prove a statement about sign changes before the M\"obius transformation, proposition \ref{prop244}.

 First we define recursively normal sequences:
\begin{lstlisting}
Fixpoint normal_seq (s : seq R) := 
   if s is (a::l1) then 
      if l1 is (b::l2) then
         if l2 is (c::l3) then 
            (normal_seq l1)
            && ((0 = a) || ((a * c $\leq$ b^2) && (0 < a) && (0 < b)))
         else (0 $\leq$ a) && (0 < b)
      else (0 < a)
   else false.
\end{lstlisting}
Then we qualify a polynomial normal, if its sequence of coefficients is normal:
\begin{lstlisting}
Definition normal := [qualify p : {poly R} | normal_seq p].
\end{lstlisting}
This definition allows us to write \lstinline!p \is normal! hereinafter.
Then we show several lemmas that guarantee that our definition of a normal polynomial agrees with the definition from section \ref{normal_math}:
\begin{lstlisting}
Lemma normal_coef_geq0 : forall (p : {poly R}),
   (p \is normal) ->
   (forall k, 0 $\leq$ p`_k). 
\end{lstlisting}

\begin{lstlisting}
Lemma normal_lead_coef_gt0 : forall (p : {poly R}),
   (p \is normal) ->
   0 < lead_coef p.
\end{lstlisting}

\begin{lstlisting}
Lemma normal_squares : forall (p : {poly R}),
   (p \is normal) ->
   (forall k, (1 $\leq$ k) -> p`_(k.-1) * p`_(k.+1) $\leq$ p`_k^2).
\end{lstlisting}

\begin{lstlisting}
Lemma normal_some_coef_gt0 : forall (p : {poly R}),
   (p \is normal) ->
   (forall i, (0 < p`_i) ->
      (forall j, (i < j)-> (j < (size p).-1) -> 0 < p`_j)).
\end{lstlisting}

\begin{lstlisting}
Lemma prop_normal : forall (p : {poly R}),
   (forall k, 0 $\leq$ p`_k) /\
   (0 < lead_coef p) /\
   (forall k, (1 $\leq$ k) -> p`_(k.-1) * p`_(k.+1) $\leq$ (p`_k)^2) /\
   (forall i, (0 < p`_i) ->
      (forall j, (i < j)-> (j < (size p).-1) -> 0 < p`_j)) ->
   p \is normal.
\end{lstlisting}
Lemma \ref{lemma241} is formalised by:
\begin{lstlisting}
Lemma monicXsubC_normal : forall (a : R),
   ('X - a) \is normal = (a $\leq$ 0).
\end{lstlisting}
The area $\ssb$ is defined by the following predicate:
\begin{lstlisting}
Definition inB (z : C) :=
   (Re z $\leq$ 0) && ((Im z)^2 <= 3 * (Re z)^2).
\end{lstlisting}
Lemma \ref{lemma242} is formalised by:
\begin{lstlisting}
Lemma quad_monic_normal : forall (z : C),
   (('X^2 - 2 * (Re z) * 'X + (Re z)^2 + (Im z)^2) \is normal) = (inB z).
\end{lstlisting}
The advantage of having formalised normal lists recursively is that in the proofs of \lstinline!monicXsubC_normal! and \lstinline!quad_monic_normal! the normal polynomials are computed by Coq automatically. 
Remark \ref{remarkno0} is formalised by the lemma:
\begin{lstlisting}
Lemma normal_0notroot : forall (p : {poly R}),
   (p \is normal) ->
   $\sim$(root p 0) $\leftrightarrow$ forall k, (k <=  (size p).-1) -> 0 < p`_k.
\end{lstlisting}
Lemma \ref{lemma243} is formalised by:
\begin{lstlisting}
Lemma normal_mulr : forall (p q : {poly R}),
   (p \is normal) ->
   (q \is normal) ->
   (p * q) \is normal.
\end{lstlisting}
Its proof is done in several steps. First we formalise a restricted version where we have additional hypotheses on \lstinline!p! and \lstinline!q!: 0 is not a root of them. Then we prove that a polynomial $P$ is normal if and only if $X^n P(X)$ is normal. Using this, one can factor out $X^{\mu_p}$ in \lstinline!p!, $X^{\mu_q}$ in \lstinline!q! and $X^{\mu_p + \mu_q}$ in their product and it suffices to show that  $pq / X^{\mu_p + \mu_q}$ is normal.

Now we can formalise Proposition \ref{prop240}:
\begin{lstlisting}
Lemma normal_root_inB : forall (p : {poly R}),
   (p \is monic) ->
   (forall z : C, root p z -> inB z) -> 
   p \is normal.
\end{lstlisting}
Its proof is similar to the one of lemma \lstinline!nonneg_root_nonpos!. The proof goes by induction on the degree of \lstinline!p!. Let $z$ be a root of \lstinline!p! so that we can factor \lstinline!p!$=(X-z)p_1(X)$. If $z$ is real, then by hypothesis $z \leq 0$ and $(X-z)$ is normal by lemma \ref{lemma241}. By the induction hypothesis $p_1$ is normal. Since the product of two normal polynomials is normal we can conclude that \lstinline!p! is normal. If $z$ has non-zero imaginary part, then $\ol{z}$ is a root too and we can factor \lstinline!p!=$(X-z)(X-\ol{z})p_2(X)$. By hypothesis $z \in \ssb$ and by symmetry $\ol{z} \in \ssb$ too. Thus $(X-z)(X-\ol{z})$ is normal by lemma \ref{lemma242}. By the induction hypothesis $p_2$ is normal. Since the product of two normal polynomials is normal, we can conclude that \lstinline!p! is normal.

Recall from subsection \ref{normal_math} proposition \ref{prop244}: it states that the number of sign changes in the sequence of coefficients of $P(X)(X-a)$ is 1, where $P$ is a normal polynomial and $a >0$. The formalised proposition is as follows:
\begin{lstlisting}
Lemma normal_changes : forall (a : R) (p : {poly R}),
   (0 < a) ->
   (p \is normal) ->
   ($\sim$(root p 0)) ->
   changes (seqn0 (p * ('X - a))) = 1.
\end{lstlisting}
The hypothesis \lstinline!$\sim$(root p 0)! is justified by remark \ref{remarkno0}.
For a better readability we introduce the notation \lstinline!n = size (p * ('X - a)).-1!. 
The formal proof of \lstinline!normal_changes! follows the ideas sketched in the proof of proposition \ref{prop244} in subsection \ref{normal_math}.

We prove first that \lstinline!(p * ('X - a))`_0 < 0! and  that \lstinline!0 < (p * ('X - a))`_n! under the same hypotheses as the ones of \lstinline!normal_changes!. 
Then we continue by distinguishing two cases concerning the length of the  sequence of coefficients of \lstinline!p * ('X - a)! (which is equivalent to distinguishing by the degree of this polynomial).

If the sequence consists of only two coefficients, then the assertion is immediately true. The sequence cannot consist of less coefficients since \lstinline!p!, which is normal, cannot be the zero polynomial. 

The main case is the one where the sequence consists of more than 2 coefficients. 
In this case we can show that the number of sign changes can be decomposed into three terms:
\begin{itemize}
 \item the number of sign changes between the first and second coefficients,
 \item the number of sign changes between the before last and last coefficients,
 \item the number of sign changes in the middle coefficients.
\end{itemize}
This decomposition is formalised by the following lemma:
\begin{lstlisting}
Lemma changes_decomp_sizegt2 : forall (s : seq R),
   (all_neq0 s) ->
   (2 < size s) ->
   changes s = (s`_0 * s`_1 < 0) + changes (mid s) +
               (s`_(size s).-2 * s`_(size s).-1 < 0)
\end{lstlisting}
The predicate \lstinline!all_neq0 s! is true iff all the items of \lstinline!s! are different from 0 and the sequence \lstinline!mid s! consists of \lstinline!s! without the first and last items. We apply \lstinline!changes_decomp_sizegt2! to \lstinline!seqn0 (p * ('X - a))!.

Next we are going to simplify the number of sign changes in the middle coefficients of \lstinline!seqn0 (p * ('X - a))!.
 Recall from subsection \ref{normal_math} that the middle coefficients of \lstinline!(p * ('X - a))! are of the form \lstinline!p`_k.+1 * (p`_k / p`_k.+1 - a)!. This sequence can be characterised as the point-wise product of the two sequences \lstinline!(drop 1 p)! and a sequence \lstinline!spseq!. This latter sequence represents the expressions \lstinline!p`_k / p`_k.+1 - a! and is formalised by
\begin{lstlisting}
Definition spseq := [seq x.1 / x.2 - a | x <- zip p (drop 1 p)].
\end{lstlisting}
The point-wise product of two sequences is formalised by \lstinline!seqmul! of type \lstinline!seq R -> seq R -> seq R!, which takes two sequences as input and returns a sequence whose items are products of the corresponding items of the two input lists. So the above mentioned characterisation is formalised by the following lemma:
\begin{lstlisting}
Lemma seqmul_spseq_dropp : mid (p * ('X -a)) = seqmul spseq (drop 1 p).
\end{lstlisting} 

Moreover we show that \lstinline!drop 1 p! consists of positive items and \lstinline!spseq! is increasing, by using the predicates \lstinline!all_pos! and \lstinline!increasing!:
\begin{lstlisting}
Lemma all_pos_dropp : all_pos (drop 1 p).
\end{lstlisting}
\begin{lstlisting}
Lemma spseq_increasing : increasing spseq.
\end{lstlisting}

 But since we apply the filter \lstinline!seqn0! on the sequence of the middle coefficients, so that \lstinline!changes! counts the sign changes correctly, we have to show some technical details due to the filter.
\begin{itemize}
 \item The filter \lstinline!seqn0! and \lstinline!mid! commute under the condition that the first and last items of a sequence are different from zero.
\begin{lstlisting}
Lemma mid_seqn0_C : forall (s : seq R),
   (s`_0 $\neq$ 0) ->
   (s`_(size s).-1 $\neq$ 0) ->
   mid (seqn0 s) = seqn0 (mid s).
\end{lstlisting}

 \item When examining closely the expressions \lstinline!p`_k.+1 * (p`_k / p`_k.+1 - a)!, we remark that 
\[p_{k+1} \biggl( \frac{p_k}{p_{k+1}} - a\biggr) = 0 \qquad \Leftrightarrow \qquad \frac{p_k}{p_{k+1}} - a = 0 \]
because $p_k >0$ for all coefficients of $P$. So the items filtered out by \lstinline!seqn0! in \lstinline!mid(p * ('X - a))! are exactly the ones filtered out by \lstinline!seqn0! in \lstinline!spseq!.
This fact is formalised by the following lemma:
\begin{lstlisting}
Lemma mid_seqn0q_decomp : mid (seqn0 (p * ('X - a))) =
   seqmul (seqn0 spseq)
          (mask [seq x != 0 | x <- mid (p * ('X - a))] (drop 1 p)).
\end{lstlisting}
Furthermore, since \lstinline!(seqn0 spseq)! is a subsequence of \lstinline!spseq!, it is increasing as well:
\begin{lstlisting}
Lemma subspseq_increasing : increasing (seqn0 spseq)
\end{lstlisting}
and since \lstinline?mask [seq x != 0 | x <- mid (p * ('X - a))] (drop 1 p)? is a subsequence of \lstinline!drop 1 p!, all its items are positive:
\begin{lstlisting}
Lemma subp_all_pos :
   all_pos (mask [seq x != 0 | x <- mid (p * ('X - a))] (drop 1 p)).
\end{lstlisting}
 \item Now we are ready to simplify the number of changes in the middle coefficients. We use the simple fact that the number of sign changes in a point-wise product of two sequences, where one of the sequences consists of positive items is equal to the number of sign changes in the other sequence. This fact is given by the lemma
\begin{lstlisting}
Lemma changes_mult : forall (s c : seq R),
   (all_pos c) ->
   (size s = size c) ->
   changes (seqmul s c) = changes s.
\end{lstlisting}
\end{itemize}
Summarising the technical lemmas, we obtain that \\ \lstinline!changes (seqn0 (mid (p * ('X - a))))! is equal to \lstinline!changes (seqn0 spseq)!.

\vspace*{0.3cm}

Just like in the proof sketch of proposition \ref{prop244} in section \ref{normal_math}, we show then that \lstinline!seqn0 spseq! has at most 1 sign change since it is increasing.

This fact is formalised for a general increasing sequence by the lemma:
\begin{lstlisting}
Lemma changes_seq_incr : forall (s : seq R),
   (increasing s) ->
   (all_neq0 s) ->
   (changes s == 1) || (changes s == 0).
\end{lstlisting} 
The conclusion of this lemma is a boolean expression, more precisely a boolean disjunction. The boolean disjunction is written \lstinline!||! in Ssreflect and the boolean equality is denoted by \lstinline!==!.

Then we proceed by distinction of cases: either 1 or 0 sign changes in \lstinline!seqn0 spseq!.
 We use the notation \lstinline!d = size (seqn0 (p * ('X - a))).-1! (and thus \lstinline!d.-1 = size (seqn0 spseq)!).
\begin{enumerate}
 \item First case : \lstinline!changes (seqn0 spseq) = 1!. This means that the first item has negative sign and the last one positive sign.
This is formalised by the lemma
\begin{lstlisting}
Lemma changes_seq_incr_1 : forall (s : seq R),
   (1 < size s) ->
   (increasing s) ->
   (all_neq0 s) ->
   (changes s == 1) = (s`_0 < 0) && (0 < s`_((size s).-1)).
\end{lstlisting}
The notations of this lemma might seem odd at first sight, since its conclusion is an equality \lstinline!=! between two expressions containing another sort of equality \lstinline!==!. The Ssreflect libraries of Coq are based on manipulating boolean expressions rather than propositions where possible. The expression \lstinline!s`_0 < 0! for example, is true or false, so is effectively a boolean. The same is valid for \lstinline!0 < s`_((size s).-1!. The operator \lstinline!&&! is the boolean conjunction. On the left-hand side of the equality the expression \lstinline!changes s == 1! is a boolean expression as well, since it uses boolean equality \lstinline!==!. So the conclusion of the lemma is an equality between boolean expressions.

Applying this lemma to \lstinline!seqn0 spseq!, we obtain \lstinline!(seqn0 spseq)`_0 < 0! on the one hand  
which implies that \\ \lstinline!(seqn0 (p * ('X - a)))`_0 * (seqn0 (p * ('X - a)))`_1 < 0! is false.

On the other hand we have 
\lstinline!0 < (seqn0 spseq)`_d.-2! which implies that 
 \lstinline!(seqn0 (p * ('X - a)))`_d.-1 * (seqn0 (p * ('X - a)))`_d < 0!
 is false.

So the count of changes according to \lstinline!changes_decomp_sizegt2!
 adds up to 1.

 \item Second case : \lstinline!changes (seqn0 spseq) = 0!.
This means that the signs of the first and last items are the same. This is formalised by the lemma:
\begin{lstlisting}
Lemma changes_seq_incr_0 : forall (s : seq R),
   (0 < size s) ->
   (increasing s) ->
   (all_neq0 s) ->
   ((changes s == 0) = (0 < s`_0 * s`_((size s).-1))).
\end{lstlisting}
Again, the assertion is an equality between boolean expressions.

So either \lstinline!0 < (seqn0 spseq)`_0! and \lstinline!0 < (seqn0 spseq)`_d.-2! or \\
\lstinline!(seqn0 spseq)`_0 < 0! and \lstinline!(seqn0 spseq)`_d.-2 < 0!. 
\\If both are positive, then 
\begin{lstlisting}
(seqn0 (p * ('X - a)))`_0 * (seqn0 (p * ('X - a)))`_1 < 0
\end{lstlisting}
is true and
\begin{lstlisting}
(seqn0 (p * ('X - a)))`_d.-1 * (seqn0 (p * ('X - a)))`_d < 0
\end{lstlisting}
is false.
If both are negative, then 
\begin{lstlisting}
(seqn0 (p * ('X - a)))`_0 * (seqn0 (p * ('X - a)))`_1 < 0
\end{lstlisting}
is false and
\begin{lstlisting}
(seqn0 (p * ('X - a)))`_d.-1 * (seqn0 (p * ('X - a)))`_d < 0
\end{lstlisting}
is true.

So the count of changes according to \lstinline!changes_decomp_sizegt2! 
 adds up to 1.
\end{enumerate}
This completes the formal proof of proposition \ref{prop244} or of \lstinline!normal_changes!, as well as the theory on normal polynomials needed for the proof of the second part.

To summarise, the theory of normal polynomials is not formalised exactly the same way the informal theory suggests. The inductive definition of normal polynomials leaves computations for Coq to conduct. The (informal) proof of lemma \ref{lemma243} is itself technical and the formal proof of \lstinline!normal_mulr! is so too, we have not found a way to avoid this. But one can proceed similarly to the informal way thanks to the Ssreflect libraries. For the formal version of proposition \ref{prop244} and its proof (which are \lstinline!normal_changes! and its proof) the filter \lstinline!seqn0! adds technical details. They appear in the simplification of \lstinline!changes (seqn0 (mid (p * ('X - a))))! to \lstinline!changes (seqn0 spseq)! and they do not arise in the informal proof.

\subsection{Second part of the proof: using normal polynomials}
\label{second_coq}

This subsection formalises the part of the proof of the second assertion of the theorem of three circles that comes after the M\"obius transformation.

First we need to formalise the union of the two disks $\ssc_1 \cup \ssc_2$. They have following equations:
\[\biggl\{x + yi \in R[i] \phantom{i} | \phantom{i} \Bigl(x - \frac{l+r}{2} \Bigr)^2 + \Bigl(y \pm \frac{\sqrt{3}(r-l)}{6}\Bigr)^2 < \frac{(r-l)^2}{3} \biggr\},\]
or equivalently
\[\Bigl\{x + yi \in R[i] \phantom{i} | \phantom{i} x^2 - (l + r)x + y^2 \pm \frac{\sqrt{3}}{3} (r-l) y+ rl < 0\Bigr\}.\]
So the union is formalised by the following predicate:
\begin{lstlisting}
Definition inC12 (l r : R) (z : C) :=
   ((Re z)^2 - (l + r) * (Re z) + (Im z)^2 - (r - l) * (Im z) / (sqrt 3) +
      l * r < 0) ||
   ((Re z)^2 - (l + r) * (Re z) + (Im z)^2 + (r - l) * (Im z) / (sqrt 3) +
      l * r < 0).
\end{lstlisting}
The second part of the theorem of three circles asserts that if $P$ has exactly one simple root in $\ssc_1 \cup \ssc_2$, then there is exactly one sign variation in the sequence of Bernstein coefficients of $P$. So in fact $P$ is of the form $P(X) = (X - a)\wt{P}(X)$ where $a \in (l,r)$ and $a$ is not a root of $\wt{P}$.
This assertion is formalised as follows:
\begin{lstlisting}
Theorem three_circles_2 : forall  (l r : R) (p : {poly R}) (a : R),
   ($\sim$(root p r)) ->
   (l < a < r) ->
   ($\sim$(root p a)) ->
   (forall z : C, root p z -> $\sim$ (inC12 l r z) ) ->
   changes (seqn0 (Mobius (p * ('X - a)) l r)) = 1.
\end{lstlisting}
The only exotic hypothesis is the one asking for $r$ not to be a root of $\wt{P}$. The reason for this is the fact that we restrict ourselves to the case that $0$ is not a root of the normal polynomial $\Mob(\wt{P})$ if want to use proposition \ref{prop244} or \lstinline!normal_changes! for the proof. To ask $0$ not to be a root of  $\Mob(\wt{P})$ is equivalent of asking for $r$ not to be a root of $\wt{P}$.

In order to apply lemma \lstinline!normal_changes!, first we need to write\\ \lstinline!Mobius(p * ('X - a))! in the form \lstinline!(Mobius p)  * ('X - b)!.
By the lemma \\ \lstinline!changes_mulC!, the multiplication of the sequence by a non-zero constant, such as the inverse of the leading coefficient of \lstinline!Mobius (p * ('X -a))!, does not affect the sign changes. 
Furthermore we show that the M\"obius transformation is compatible with the product of polynomials:
\begin{lstlisting}
Lemma MobiusM : forall (p q : {poly R}) (l r : R),
   Mobius (p * q) l r = (Mobius p l r) * (Mobius q l r).
\end{lstlisting}
We can compute explicitly the coefficients of the M\"obius transformation of a monic polynomial of degree 1:
\begin{lstlisting}
Lemma Mobius_Xsubc_monic : forall (a l r : R),
   (l != r) ->
   (l != a) ->
   (lead_coef (Mobius ('X - a) l r))^(-1) * (Mobius ('X - a) l r) 
         = 'X + ((r - a) / (l - a)).
\end{lstlisting}
Now we can apply lemma \lstinline!normal_changes! and it remains to show its hypotheses.
The first hypothesis $\frac{r-a}{l-a} < 0$ can be shown easily since $l < a <r$.

To show the hypothesis that 
\begin{lstlisting}
(lead_coef (Mobius p l r))^(-1) * (Mobius p l r) \is normal,
\end{lstlisting}
 we apply lemma \lstinline!normal_root_inB!. This polynomial is obviously monic. In order to show that all the roots of \lstinline!(Mobius p l r)! are in $\ssb$, we show that it is equivalent to ask that all the roots of \lstinline!p! are in the exterior of $\ssc_1 \cup \ssc_2$. We use thus the following lemma
\begin{lstlisting}
Lemma inB_notinC12 : forall (l r : R) (z : C),
   (l != r) ->
   (z + 1 $\neq$ 0) ->
   (inB z) = $\sim \sim$(inC12 l r ((r + l * z) / (z + 1)))
\end{lstlisting}
together with lemma \lstinline!root_Mobius_C_2! from section \ref{first_coq}. The conclusion of lemma \lstinline!inB_notinC12! is an equality of boolean expressions using the boolean negation, denoted by $\sim\sim$.

So we have obtained the equivalence between ``all roots of  \lstinline!(Mobius p l r)! are in $\ssb$" and ``all roots of \lstinline!p! are in the exterior of $\ssc_1 \cup \ssc_2$".

What happens here is intuitively similar to section \ref{first_coq}. We keep track of the complementary area of $\ssc_1 \cup \ssc_2$ when doing the M\"obius transformation: translation by $-l$, then scaling by $r-l$, inversion and translation by $-1$. 

The computations are not as immediate as in section \ref{first_coq}, but lemma \lstinline!inB_notinC12! proves the correctness. The case of a root in $z = -1$ has to be treated apart, but without any further difficulty. This concludes the formal proof of the second assertion.

This part of the proof can be formalised the way that the informal counterpart suggests.

\section{Discussion and future work}
\label{future}

First we would like to mention some technical remarks concerning the formalisation the way it was carried out.

We implemented normal polynomials recursively because in the proofs of \lstinline!monicXsubC_normal! and \lstinline!quad_monic_normal! the computations are carried out automatically by Coq. Whereas in an alternative definition by a (rather cumbersome) predicate, one would have to show ``by hand" the four defining properties. Another simplification of normal polynomials, considering remark \ref{remarkno0}, would have been to formalise only normal polynomials without any root in 0, since in almost all proofs thereafter, we add this hypothesis. We consider this simplification as part of our future work.

Another possibility of improving the proof is to develop apart a (small) theory of the function \lstinline!fun s : seq R => changes (seqn0 s)! counting the sign changes in a sequence discarding the 0 items. It would make the proof more elegant, and the technical details such as the ones needed for the proof of \lstinline!normal_changes! woud be treated in this theory and sourced out from the proof of the theorem of three circles.

As mentioned in the introduction, one of our motivations for formalising the theorem of three circles is to provide the main pieces for the formal proof of the termination of the algorithm of real root isolation as described in \cite{bpr}. A possible future work would be to actually put the pieces together and formalise the algorith itself and its termination.

Another one of our motivations for formalising the theorem of three circles is to formalise the general case involving Obreshkoff areas and lenses. Chapter two of \cite{eigenwillig_phd} explains all the details, presenting the relevant works of Obreshkoff. Lemmas \lstinline!nonneg_changes0! and \ref{prop244}, or its formalised version \lstinline!normal_changes!, are generalised by the following theorem of Obreshkoff  (restated in \cite{eigenwillig_phd} as Theorem 2.7)
\begin{mtheorem}[Obreshkoff]
Consider the real polynomial $P(X)=\sum_{i=0}^n a_i X^i$ of degree $n$ and its complex roots, counted with multiplicities. Let $v$ denote the number of sign changes of the sequence $(a_0, \ldots, a_n)$. If $P(X)$ has at least $p$ roots with arguments in the range $-\frac{\pi}{n+2-p} < \varphi < \frac{\pi}{n+2-p}$, and at least $n-q$ roots with arguments in the range $\pi - \frac{\pi}{q+2} \leq \psi \leq \pi + \frac{\pi}{q+2}$, then $p \leq v \leq p$. If $p=q$, then $P(X)$ has exactly $p$ roots with arguments $\varphi$ in the range given above and $v = p$.
\label{theorem227}
\end{mtheorem}

The special case $p = q = 0$ is our lemma \lstinline!noneg_changes0!:
the range for $\psi$ is $[\frac{\pi}{2};\frac{3\pi}{2}]$, which corresponds to roots with non-positive real part.
 
The special case $p=q=1$ is our lemma \ref{prop244} or \lstinline!normal_changes!: the range for $\psi$ is $[\frac{2\pi}{3};\frac{4\pi}{3}]$ which corresponds to the area $\ssb$ and one complex root (without its complex conjugate) with argument in the range $(-\frac{\pi}{n+1}; \frac{\pi}{n+1})$ implies that this root is in fact real and positive.

The transformation of polynomials from proposition \ref{bernQ} in order to obtain the M\"obius transform or \lstinline!Mobius p! is characterised in \cite{eigenwillig_phd} as the M\"obius transformation $I$ of the interval $(0,\infty)$ to an arbitrary open interval $(l,r)$:
\[P(X) \mapsto (X+1)^n P\biggl(\frac{r + lX}{X+1}\biggr) \]
which is the transformation we use in lemma \lstinline!root_Mobius_C_2! (as well as in \lstinline!notinC_Re_lt0_2! and \lstinline!inB_notinC12!). This is our reason for calling the sequence of the four transformations in proposition \ref{bernQ} a M\"obius transformation.

The generalisation of the theorem of three circles is obtained by transferring theorem \ref{theorem227} to an arbitrary interval $(l,r)$. To do so we define first Obreshkoff areas and lenses.

The two Obreshkoff discs $\ol{\ssc}_k$ and $\ul{\ssc}_k$ for an integer $k$ are the open discs whose delimiting circles touch the endpoints of $(l,r)$ and whose centers see the line segment $(l,r)$ under the angle $\frac{2\pi}{k+2}$.
The Obreshkoff area $A_k$ is the union $\ol{\ssc}_k \cup \ul{\ssc}_k$ and the Obreshkoff lens $L_k$ is the intersection $\ol{\ssc}_k \cap \ul{\ssc}_k$.

\begin{mtheorem}[Obreshkoff]
Consider the real polynomial $P(X)$ of degree $n$ and its roots in the complex plane, counted with multiplicities. Let $v$ denote the number of sign changes in the sequence of coefficients of the M\"obius transformation of $P$ to the interval $(l,r)$.
If $P$ has at least $p$ roots in the Oreshkoff lens $L_{n-p}$ and at most $q$ roots in the Obreshkoff area $A_q$, then $p \leq v \leq q$.

\label{theorem232}
\end{mtheorem}
The assertions of the theorem of three circles are the special cases of $p=q=0$ and $p=q=1$.

In order to formalise these two theorems \ref{theorem227} and \ref{theorem232}, one would have to make the following changes (at least) in the definitions of the structures needed for the proofs.
\begin{itemize}
 \item Adapt the definition of the discs $\ssc_1$ and $\ssc_2$ to Obreshkoff areas with at least one integer parameter. Formalise Obreshkoff lenses.
 \item Adapt the definition of the area $\ssb$, with at least one more parameter, define the area corresponding to the angle range for $\varphi$.
 \item Adapt the definition of normal polynomial with respect to the pa\-ra\-met\-rised version of $\ssb$. If possible keep recursive definition, in order to make computations automatic when possible.
\end{itemize}

To carry out the proof of theorem \ref{theorem227}, one would have to prove first the special case $q=n$. This proof could be done in two steps: first an analogous lemma to \lstinline!normal_root_inB!, with a proof by induction on the degree of the polynomial $P$, and then an analogous one to \lstinline!normal_changes! which provides an upper bound for the number of sign changes rather than an exact number of changes. Then with this special case one can prove the general statement of theorem \ref{theorem227} by applying the special case to $P(X)$ and $P(-X)$.

To prove theorem \ref{theorem232}, we need first the mentioned M\"obius transformation. For this purpose lemma \lstinline!root_Mobius_C_2! together with a similar lemma to \lstinline!inB_notinC12! are mostly enough. The proof of theorem \ref{theorem232} would be then quite analogous to the one of \lstinline!three_circles_2!.

Bearing in mind that our work provides tools for the formalisation of correctness of the NewDsc algorithm for example, we would have accomplished the first step in this direction. But until the completion of this goal, we still have many opportunities for exploration of formalisation of mathematical theories.

\bibliographystyle{spmpsci}
\bibliography{bib_3circles}

\end{document}